\documentclass[fleqn,10pt]{wlscirep}
\usepackage[utf8]{inputenc}
\usepackage[T1]{fontenc}
\usepackage{lineno}
\usepackage{url}
\usepackage{listings}

\title{Mosqlimate: a platform to providing automatable access to data and forecasting models for arbovirus disease.}

\author[1]{Fabiana Ganem}
\author[1]{Luã Bida Vacaro}
\author[1]{Eduardo Correa Araujo}
\author[2,3]{Leon Diniz Alves}
\author[2]{Leonardo Bastos }
\author[1]{Luiz Max Carvalho}
\author[2]{Iasmim Almeida}
\author[4]{Asla Medeiros de Sá}
\author[1,*]{Flávio Codeço Coelho}
\affil[1]{Fundação Getulio Vargas, Applied Mathematics School, Rio de Janeiro, 22250-900, Brazil}
\affil[2]{Fundação Oswaldo Cruz, Programa de Computação Científica, Rio de Janeiro, Brazil}
\affil[3]{Centro Federal de Educação Tecnológica Celso Suckow da Fonseca, Rio de Janeiro, 20271-204, 
Brazil}
\affil[4]{Instituto de Matemática Pura e Aplicada - IMPA Tech - Rio de Janeiro, 20220-800 , Brazil} 
\affil[*]{corresponding author(s): Flávio Codeço Coelho (fccoelho@fgv.br)}

\begin{abstract}
Dengue is a climate-sensitive mosquito-borne disease with a complex transmission dynamic. Data related to climate, environmental and sociodemographic characteristics of the target population are important for project scenarios. Different datasets and methodologies have been applied to build complex models for dengue forecast, stressing the need to evaluate these models and their relative accuracy grounded on a reproducible methodology. The goal of this work is to describe and present Mosqlimate, a web-based platform composed by a dashboard, a data store, model and rediction registries and support for a community of practice in arbovirus forecasting. Multiple API endpoints give access to data for development, open registration of predictive models from different approaches and sharing of predictive models for arboviruses incidence, facilitating interaction between modellers and allowing for proper comparison of the performance of different registered models, by means of probabilistic scores. Epidemiological, entomological, climatic and sociodemographic datasets related to arboviruses in Brazil, are freely available for download, alongside full documentation.

\end{abstract}
\begin{document}

\flushbottom
\maketitle

\thispagestyle{empty}
\section*{Background \& Summary}

Dengue is a climate-sensitive, vector-borne disease that challenges health surveillance systems worldwide.
It is a growing global public health threat, requiring detailed observational and experimental data to help understand its determinants and complex mode of transmission.
These are crucial to improve detection, diagnosis, vector mitigation and prevention~\cite{brady_refining_2012}.
For short-term dengue forecasting, factors related to climate and environmental characteristics seem to be more relevant while, sociodemographic characteristic seems to be more important for medium to long-term scenarios\cite{zhao_machine_2020, mussumeci2020large}.  

\textit{Aedes} mosquito populations are very sensitive to environmental factors and meteorological conditions, such as temperature, humidity and precipitation, which directly influence the spread of dengue fever\cite{xu_forecast_2020,alves2021framework}.
Previous studies have identified an association between climate and dengue outbreaks \cite{salim_prediction_2021}, and also by the increase in transmission and expansion of the disease in regions currently free of dengue such as Europe, Asia, North America and Australia\cite{sebastianelli_reproducible_2024}.

The literature on arbovirus disease forecasting is quite extensive, including a broad range of models to dengue forecast, most of them focusing on disease incidence\cite{roster_machine-learning-based_2022,nguyen_deep_2022,johansson_evaluating_2016, zhao_machine_2020, withanage_forecasting_2018,xu_forecast_2020, polwiang_time_2020, sebastianelli_reproducible_2024,sylvestre_data-driven_2022,alves2021framework}.  Mussumeci and Coelho\cite{mussumeci2020large} evaluated the potential of random forest (RF) machine learning (ML) algorithm and artificial neural network (ANN), showing that long short-term memory (LSTM) model enabled more efficient dengue incidence weekly predictions using meteorological data and dengue case compared with other candidate models, also demonstrating the importance of including sociodemographic data as predictors.
Ensembles of different models, such as neural networks, support vector regression and linear regression models, reduced prediction errors providing greater accuracy.
A study with data of Brazil and Peru, propose a machine learning ensemble model for forecasting the dengue incidence rate by identifying factors triggering dengue outbreaks, dealing with data scarcity and operational scalability challenges\cite{shashvat_ensemble_2019, sebastianelli_reproducible_2024}.

Studies focused on comparing and evaluating models and datasets for early forecast have been previously presented in the literature\cite{roster_machine-learning-based_2022},  assessing the performance of real-time forecast using reported data from epidemiological surveillance,  \cite{reich_challenges_2016}, comparing the accuracy of different mathematical approaches to predict dengue cases\cite{polwiang_time_2020}, or finally evaluating  deep learning methods employed for long- and short-term forecast \cite{nguyen_deep_2022}.
A systematic review identified several algorithms and different approaches to monitor and predict dengue outbreaks or dengue-related outcomes, identifying the most significant predictors, concluded that the main predictive models used were extremely related to the  outcome of interest  \cite{sylvestre_data-driven_2022}. 

Due to variations in the datasets used by different disease prediction studies,  model comparing can be a substantively complex task\cite{Johansson_Reich_Hota_Brownstein_Santillana_2016}.
Many datasets used for modeling may not be openly accessible to other modelers to incorporate into their own analyses, making it harder to assess the strengths and weaknesses of each model with regard to accuracy, sensitivity, specificity, and precision\cite{salim_prediction_2021}. 

The goal of this work is to describe and present the Mosqlimate platform, an API created to facilitate the development, registration  and sharing of predictive models for arboviruses, such as dengue, chikungunya and zika, providing automated access to different sets of epidemiological, sociodemographic and climatic data from Brazil, adaptable to other locations.
The goal is to allow interaction within a community of modelers through a dashboard that facilitates the comparison between the performance and results of different registered  models, facilitating the research and evaluation of the impact of climate on the dynamics of vector-borne diseases in the face of climate change. 

\section*{Methods}

The Mosqlimate platform \cite{Mosqlimate-project,Vacaro_Coelho_Araujo_Loch_Bot_Pimpale_2024} is a web-based platform comprised of an open data repository acessible through a REST API and aims to provide automated access to several arbovirus-related epidemiological, entomological, climate and sociodemographic open datasets,  previously tested, evaluated and documented.
The plataform also provides an API for open registration of predictive models, from different approaches,  for arbovirus disease incidence.
The platform also provides visualization tools to facilitate model comparison as long as their code is open and available through a public code repository, making them visible and public to the community.

\subsection*{Architecture}
\begin{figure}[ht]
    \centering
    \includegraphics[width=0.55\linewidth]{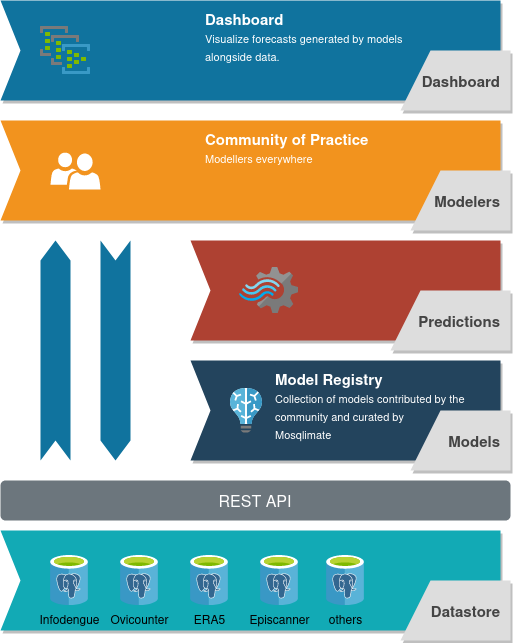}
    \caption{Data platform architecture. The four main layers are show here. the Model registry and the predictions form a single layer.}
    \label{fig:arch}
\end{figure}

The platform is composed by the following  layers: a dashboard, a data store,  model and predictions registry,  and a community of practices.
It was used a Web service conforming to a RESTful API architecture serving local and remotely hosted datasets.
The dashboard can display the forecasts produced by the models registered in the platform against observational data.
The overall structure of the platform is shown in figure \ref{fig:arch}.


The datastore layer provides multiple datasets relevant for arbovirus modeling, stored locally on the platform's PostgreSQL server or accessed remotely. The two data platforms, Infodengue (\url{https://info.dengue.mat.br/}) and Ovicounter (\url{https://contaovos.com/pt-br/}) are examples of datasets accessed remotely.
Lastly, the usage of this API is facilitated by interactive pages that help modelers build the request URLs appropriate to their data needs. 

The REST API layer unifies the programmatic interface for accessing all datasets, in a way that is uniform and well documented.
The API offers human-readable\cite{Mosqlimate-docs} documentation and automatic and interactive documentation following the OPENAPI standards\cite{OpenAPI-Initiative}.

The model registry component is designed to catalog and store metadata about forecasting models developed by Mosqlimate's community of practice. It focuses on models that provide open source implementations. The Inclusion of models in the registry is also done through the API. 

The prediction store component (red block in figure \ref{fig:arch}) is integrated with the model registry, since only registered models can upload predictions to the platform. Prediction uploads are also done through the API and its operation is detailed below.

The Community of practice, built around the platform, represents the users (modellers and public health professionals) of the platform and its developers, which interact to continue to improve the platform for their own benefit.

The most immediately visible component of the platform, the Dashboard, is designed to  allow for the visual comparison of predictions generated by different models. A detailed description of it is given below.

The Mosqlimate platform is open to anyone who wants to browse its content without the need to register. For users who want to contribute models and predictions, it is necessary to have a GitHub account to register and obtain an API KEY, thus, the full leverage of the platform becomes possible.  

\subsection*{Model classification and annotation}

When we consider the variety of disease forecasting models available in the literature\cite{Chaves_Pascual_2007,Fu_Chen_Dong_Luo_Miao_Song_Huang_Sun_2019,Johansson_Reich_Hota_Brownstein_Santillana_2016}, it becomes clear that if we want to be able to compare them in terms of forecasting skill, for example, we first need to place them in consistent categories that can be properly compared. Such a categorization task can be based on different aspects of the model. For this application, we chose to consider the following  categories: the scope of their outputs, for example space and time ranges and resolutions, and the type of the output (quantitative or qualitative). These categories are in no way exhaustive, but cover an ample set of possible models of epidemiological interest.


When registering models on the platform, the user chooses a category and provides additional information that will be stored together with the model as metadata, this makes it easier to retrieve it from the model database and also to group it with comparable models. The required metadata is described in the platform documentation as a guide for contributors, and the metadata fields that are required for all models registered on the platform are listed in the \ref{tab:model_metadata} table. These fields can also be used to filter models registered on the platform.


The \textit{Name} and \textit{Description} fields, in the model's metadata, are free textual information contributed by the author to help identify the model. The \textit{Implementation language} must be chosen from a list of languages available. This field is to help other modelers find models implemented in familiar languages. \textit{Type} is one of the categories listed in table \ref{tab:model_metadata}. \textit{Adm Level} refers to the smallest spatial scale of the model output: city(ADM2), state(ADM1) or national(ADM0), for example. Lastly, the \textit{Time resolution} field is one of daily, weekly, monthly or yearly and refers to the output of the model.

\subsection*{Storing Forecasts}
The execution and the uploading of the Models and it's Predictions' results are done by the Model's authors, off-platform, this way the platform does not need to collect information on the software dependencies of each Model. The Model's repository must be public, hosted in GitHub or GitLab; the version control (commit ID) has to be referenced in the Prediction's metadata at upload, enabling other modelers to reproduce the same results in their local machines. To upload a Model's information and its Predictions' results into the platform, the API accepts JSON objects with its required fields. The fields associated with Models and Predictions are listed in the tables \ref{tab:model_metadata} and \ref{tab:pred_metadata}, respectively. In the Predictions' metadata, \textit{model} is a reference to a Model registered in the platform; \textit{description} is an open text field where the modeler can add relevant comments about the prediction; \textit{commit} is a git hash number referring to the specific version of the Model that generated the Prediction in the git repository; \textit{predict\_date} is the date when the prediction was generated; lastly, \textit{prediction} is the output of the model, also as a JSON object.


The \textit{prediction} object posted must contain the columns presented in Table \ref{tab:pred_table_metadata}. It is mandatory to fill the  `adm\_N`  column that represents the ADM level of the model metadata (see Table \ref{tab:model_metadata}) with valid non-null values, as well as the columns: \textit{date}, \textit{pred}, \textit{upper} and \textit{lower}. If the model associated with the prediction registered has an ADM level 1, then the predictions must contain the `adm\_1` column, while the other ADM columns are not required.


Mosqlient is a Python API created to facilitate the use of the API (\url{https://github.com/Mosqlimate-project/mosqlimate-client}). In this package, there is a function called \verb|upload_prediction| that can be used to upload predictions to the platform. The function accepts several parameters as input, described in its documentation, the client allow,  the parameter \verb|prediction|, either a JSON String, a dictionary or a DataFrame containing specific columns.

\subsection*{Model Comparison}\label{sec:model_comparison}
For comparing the forecast performance of the models registered in the platform, we focused on adopting proper statistical scoring rules\cite{Gneiting_Raftery_2007}. 

To compare the predictions stored in the platform since the users provide lower, upper and median predicted values, following \cite{johansson2019open, Gneiting_Raftery_2007,bracher2021evaluating} were implemented the CRPS (Continuous Ranked Probabilistic Score) and Logarithmic Score. The implementation used the `scoringrules` Python package described in \cite{zanetta_scoringrules_2024}. The implementation assumes that the predictive distribution of the models follows a Gaussian distribution. In this way, following \cite{zanetta_scoringrules_2024}, the CRPS was computed using the equation \eqref{eq:normal_crps}: 
\begin{equation}
\label{eq:normal_crps}
    \operatorname{CRPS}(\mathcal{N}(\mu,\sigma), y) = \sigma \left\{ \omega[\Phi(\omega) - 1] + 2\phi(\omega) - \frac{1}{\sqrt{\pi}}\right\},
\end{equation}
where \( \Phi(\omega) \) and \( \phi(\omega) \) represent the cumulative distribution function (CDF) and the probability density function (PDF) of the standard normal distribution, respectively, evaluated at the normalized prediction error \( \omega = \frac{y - \mu}{\sigma} \). In the predictions saved on the platform $\mu$ stands for the preds columns and $\sigma$ is computed assuming a 95\% prediction interval, in this way, $\sigma=\frac{u-l}{4}$, being $u$ the upper column and $l$ the lower column. Analogous, the logarithmic score is computed using the equation
\eqref{eq:normal_log}: 
\begin{equation}
\label{eq:normal_log}
    \operatorname{LogS}(\mathcal{N}(\mu,\sigma), y) = \log \left(\cfrac{\phi(\omega)}{\sigma}\right).
\end{equation}
In addition to the proper scoring rules, point metrics such as MSE (Mean Squared Error) and MAE (Mean Absolute Error) are also implemented. All the metrics are available in the `mosqlient` Python package and can be seen in the dashboard created to visualise the predictions in the platform. 

\section*{Data Records}
The Mosqlimate platform datastore, gives access to Disease notification data, climate time series, such as temperature and precipitation, epidemiological statistics, and mosquito abundance data. Currently, the geographical scope of the API is Brazil, at the municipality level.

\subsection*{Infodengue data}
This is one of the key datasets made available on the platform, that consists of weekly counts of dengue, zika, and chikungunya cases in all Brazilian cities with reported cases since 2010. The Infodengue dataset\cite{codeco2018infodengue} contains a rich set of epidemiological variables that are described in the API (\url{https://info.dengue.mat.br/services/api}) and presented in the table \ref{tab:infodengue}.


\subsection*{Episcanner data}
The episcanner tool produces a unique dataset of epidemiological statistics about epidemic years in Brazil at the level of the municipality, since 2010\cite{araujo2024large}. The Episcanner dataset variable dictionary is given in table \ref{tab:episcanner-dict}.


\subsection*{Climate data}
Climate time series for all Brazilian municipalities at daily time resolutions are regularly extracted from the Copernicus ERA5 reanalysis dataset \cite{hersbach2018era5}.
The horizontal resolution of this dataset is $0.25$ degrees latitude and longitude. Hourly data are calculated for each municipality with minimum, maximum and mean values retained for each day. 
The variables available in this dataset are described in the mosqlimate API (\url{https://api.mosqlimate.org/docs/datastore/GET/climate/}) and presented in table \ref{tab:clim_vars}.


\subsection*{Ovitraps data with automated Egg Count}

The "Conta Ovos" project, in english "egg counter", represents a collaborative tool between Brazilian state and municipal governments, the Oswaldo Cruz Foundation (Fiocruz) and the Mosqlimate project (\url{https://contaovos.com/en-us/}).
This project harnesses the potential of ovitraps to monitor and control dengue fever by gathering and analyzing data on the prevalence of \textit{Aedes aegypti} mosquito eggs.
The program includes an educational component that explains the significance of ovitraps in dengue surveillance, complemented by the "Conta Ovos" mobile app, that serves as a critical tool for data collection and analysis.

The mobile app allows health surveillance agents to record and upload data on the number of eggs per ovitrap. The app automatically generates detailed maps pinpointing each ovitrap's location along with the associated egg count data. Additionally, the app enables users to download comprehensive data reports, including the status of each ovitrap (positive or negative for mosquito eggs), exact latitude and longitude, installation date, epidemiological week, and year.

The app facilitates the creation of a dynamic database that municipal health departments can use to identify high-risk areas, characterized by high concentrations of \textit{Aedes aegypti }eggs.
This data-driven approach empowers municipalities to deploy health agents more strategically, focusing on areas most in need of interventions to reduce mosquito breeding sites.
By providing real-time data and historical trends of mosquito egg counts, the project aids in crafting more responsive and informed strategies for dengue control.
The collaborative nature of the project, involving various governmental and research institutions, enhances its scalability and adaptability to different regions.

All data generated by the "Conta ovos" project are accessible through Mosqlimate, a valuable resource for researchers focusing on vector-borne diseases like dengue.
This dataset, which tracks the population dynamics of \textit{Aedes aegypti} mosquitoes through egg counts from ovitraps, offers a direct indicator of potential dengue outbreaks.
By analyzing these data, scientists can develop predictive models to forecast dengue cases more accurately.
The comprehensive data available on Mosqlimate includes geographic coordinates, egg counts, installation dates, and status updates of each ovitrap, providing a robust foundation for epidemiological studies and the development of targeted dengue control strategies.

\subsection*{\textit{Aedes aegypti} Egg Image Dataset}

The Mosqlimate Aedes aegypti Egg Image Dataset represents a valuable resource for researchers, interested in the surveillance of\textit{ Aedes} mosquitoes populations.
It consists of 700 meticulously curated JPEG images offering a comprehensive collection of wooden pallets containing \textit{Aedes aegypti} eggs.
It provides a broad dataset to train models for automated egg counting. 

The data collection process was conducted with precision and attention to detail, ensuring the integrity and accuracy of the images.
Each photo captures a particular substrate containing \textit{Aedes aegypti} eggs, with accompanying metadata stored in a CSV file.
This metadata includes: pallet ID, installation day, file type (JPEG, PNG…), flash use (Yes or No), dimensions of the photo (height and width), cellphone used to take the picture, number of eggs counted in the laboratory, material (Eucatex, White Cotton or Brown Cotton), link to the picture, and file name.
The dataset is the result of collaborative efforts within the Mosqlimate group and Fundação Oswaldo Cruz Entomology Laboratory. 

The potential applications of the Dataset are vast and varied.
Primarily, it serves as an essential resource for the development and validation of automated egg-counting applications aimed at accurately counting \textit{Aedes aegypti} eggs.
By leveraging machine learning and computer vision techniques, researchers can train algorithms to detect and count eggs within the provided images, thereby streamlining surveillance and control efforts. 

Furthermore, the dataset encourages interdisciplinary research endeavors, inviting collaborations between experts in entomology, epidemiology, data science, and public health. Beyond its immediate utility in automated counting applications, the dataset holds promise for exploring broader research questions related to mosquito population dynamics, disease transmission, and the efficacy of control measures.

\section*{Technical Validation}


In this section, we present the output of a few forecasting models stored on our platform. Their forecasts are entirely based on data available in the platform, as required by our terms of service to modelers.

\begin{figure}[h]
    \centering
    \includegraphics[width=0.8\linewidth]{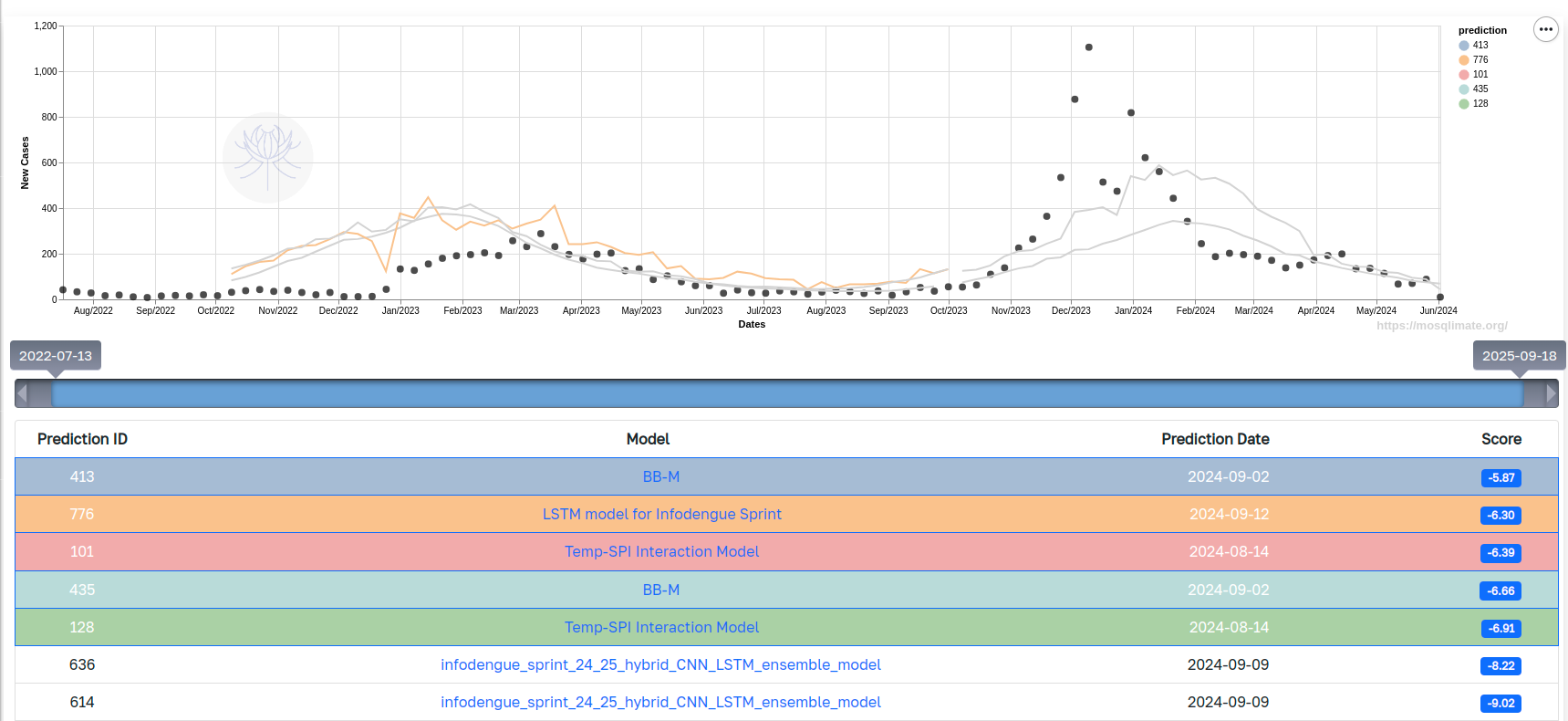}
    \caption{Predictions registered in the platform in the state of Minas Gerais between 2022-10-09 and 2023-09-24.}
    \label{fig:example_sprint}
\end{figure}

In Figure \ref{fig:example_sprint}, there is an example of how the predictions registered in the platform can be visualized. The solid lines represent the predictions, and the black points the real data. Hovering the mouse over a line in the left panel will update the right panel, showing the predictive interval for the selected model. The colors of the charts and the rows of the table above are matched, to facilitate identifying the selected model's metadata.

\begin{figure}[h]
    \centering
    \includegraphics[width=0.65\linewidth]{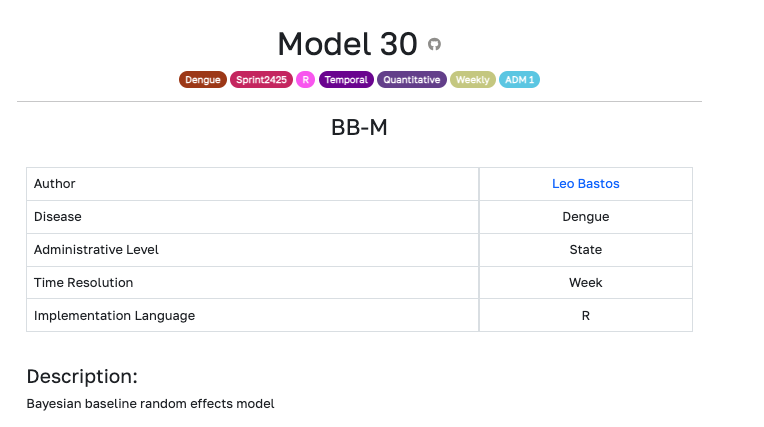}
    \caption{Description page of the Model 'BB-M'.}
    \label{fig:desc_model}
\end{figure}

The column `Model' of the table contains the link to the model record in the platform that provides some information about the model, as shown in Figure \ref{fig:desc_model}. The column \verb|Score| actually is a dropdown menu that allows the user to compare the prediction according to one of the following scores: MAE: Mean Absolute Error, MSE: Mean Squared Error, 'log\_score': Logarithmic Score and CRPS: Continuous Ranked Probability Score. The scores are computed as described in the methods section.


\section*{Usage Notes}

The Mosqlimate data platform data was designed to be accessible both manually and automatically by third-party code. The REST API is language agnostic, requiring only that developers know how to program HTTP requests to the platform. Automation examples are available in the documentation for the \verb|Python| and \verb|R| programming languages, as well as for direct access from the Linux terminal using the \verb|curl| program. Listing \ref{lst:dengue} shows a quick example of how to fetch dengue incidence data. The platform only makes available data that is already in the public domain.

\begin{lstlisting}[language=python, caption=Example of fetching dengue incidence data using Python., label=lst:dengue]
import requests

infodengue_api = "https://api.mosqlimate.org/api/datastore/infodengue/"

page = 1
pagination = f"?page={page}&per_page=100&"
filters = "disease=%s&start=%s&end=%s" % ("dengue", "2022-12-30", "2023-12-30")

resp = requests.get(infodengue_api + pagination + filters) # GET request

items = resp.json()["items"] # JSON data in dict format
resp.json()["pagination"] # Pagination*
\end{lstlisting}

\section*{Code availability}

All the code related to the implementation of the data platform is available on the Mosqlimate project GitHub repositories\cite{Mosqlimate-project}.

Instructions on how to use the platform are also available online~\cite{Mosqlimate-docs}.
\bibliography{refs}

\begin{thebibliography}{10}
\urlstyle{rm}
\expandafter\ifx\csname url\endcsname\relax
  \def\url#1{\texttt{#1}}\fi
\expandafter\ifx\csname urlprefix\endcsname\relax\def\urlprefix{URL }\fi
\expandafter\ifx\csname doiprefix\endcsname\relax\def\doiprefix{DOI: }\fi
\providecommand{\bibinfo}[2]{#2}
\providecommand{\eprint}[2][]{\url{#2}}

\bibitem{brady_refining_2012}
\bibinfo{author}{Brady, O.~J.} \emph{et~al.}
\newblock \bibinfo{journal}{\bibinfo{title}{Refining the {Global} {Spatial}
  {Limits} of {Dengue} {Virus} {Transmission} by {Evidence}-{Based}
  {Consensus}}}.
\newblock {\emph{\JournalTitle{PLoS Neglected Tropical Diseases}}}
  \textbf{\bibinfo{volume}{6}}, \bibinfo{pages}{e1760},
  \url{10.1371/journal.pntd.0001760} (\bibinfo{year}{2012}).

\bibitem{zhao_machine_2020}
\bibinfo{author}{Zhao, N.} \emph{et~al.}
\newblock \bibinfo{journal}{\bibinfo{title}{Machine learning and dengue
  forecasting: {Comparing} random forests and artificial neural networks for
  predicting dengue burden at national and sub-national scales in {Colombia}.}}
\newblock {\emph{\JournalTitle{PLoS Negl Trop Dis}}}
  \textbf{\bibinfo{volume}{14}}, \bibinfo{pages}{e0008056},
  \url{10.1371/journal.pntd.0008056} (\bibinfo{year}{2020}).
\newblock \bibinfo{note}{Place: United States}.

\bibitem{mussumeci2020large}
\bibinfo{author}{Mussumeci, E.} \& \bibinfo{author}{Coelho, F.~C.}
\newblock \bibinfo{journal}{\bibinfo{title}{Large-scale multivariate
  forecasting models for dengue-lstm versus random forest regression}}.
\newblock {\emph{\JournalTitle{Spatial and Spatio-temporal Epidemiology}}}
  \textbf{\bibinfo{volume}{35}}, \bibinfo{pages}{100372}
  (\bibinfo{year}{2020}).

\bibitem{xu_forecast_2020}
\bibinfo{author}{Xu, J.} \emph{et~al.}
\newblock \bibinfo{journal}{\bibinfo{title}{Forecast of {Dengue} {Cases} in 20
  {Chinese} {Cities} {Based} on the {Deep} {Learning} {Method}.}}
\newblock {\emph{\JournalTitle{Int J Environ Res Public Health}}}
  \textbf{\bibinfo{volume}{17}}, \url{10.3390/ijerph17020453}
  (\bibinfo{year}{2020}).
\newblock \bibinfo{note}{Place: Switzerland}.

\bibitem{alves2021framework}
\bibinfo{author}{Alves, L.~D.}, \bibinfo{author}{Lana, R.~M.} \&
  \bibinfo{author}{Coelho, F.~C.}
\newblock \bibinfo{journal}{\bibinfo{title}{A framework for weather-driven
  dengue virus transmission dynamics in different brazilian regions}}.
\newblock {\emph{\JournalTitle{International Journal of Environmental Research
  and Public Health}}} \textbf{\bibinfo{volume}{18}}, \bibinfo{pages}{9493}
  (\bibinfo{year}{2021}).

\bibitem{salim_prediction_2021}
\bibinfo{author}{Salim, N. A.~M.} \emph{et~al.}
\newblock \bibinfo{journal}{\bibinfo{title}{Prediction of dengue outbreak in
  {Selangor} {Malaysia} using machine learning techniques.}}
\newblock {\emph{\JournalTitle{Sci Rep}}} \textbf{\bibinfo{volume}{11}},
  \bibinfo{pages}{939}, \url{10.1038/s41598-020-79193-2}
  (\bibinfo{year}{2021}).
\newblock \bibinfo{note}{Place: England}.

\bibitem{sebastianelli_reproducible_2024}
\bibinfo{author}{Sebastianelli, A.} \emph{et~al.}
\newblock \bibinfo{journal}{\bibinfo{title}{A reproducible ensemble machine
  learning approach to forecast dengue outbreaks.}}
\newblock {\emph{\JournalTitle{Sci Rep}}} \textbf{\bibinfo{volume}{14}},
  \bibinfo{pages}{3807}, \url{10.1038/s41598-024-52796-9}
  (\bibinfo{year}{2024}).
\newblock \bibinfo{note}{Place: England}.

\bibitem{roster_machine-learning-based_2022}
\bibinfo{author}{Roster, K.}, \bibinfo{author}{Connaughton, C.} \&
  \bibinfo{author}{Rodrigues, F.~A.}
\newblock \bibinfo{journal}{\bibinfo{title}{Machine-{Learning}-{Based}
  {Forecasting} of {Dengue} {Fever} in {Brazilian} {Cities} {Using}
  {Epidemiologic} and {Meteorological} {Variables}.}}
\newblock {\emph{\JournalTitle{American journal of epidemiology}}}
  \textbf{\bibinfo{volume}{191}}, \bibinfo{pages}{1803--1812},
  \url{10.1093/aje/kwac090} (\bibinfo{year}{2022}).
\newblock \bibinfo{note}{Place: United States}.

\bibitem{nguyen_deep_2022}
\bibinfo{author}{Nguyen, V.-H.} \emph{et~al.}
\newblock \bibinfo{journal}{\bibinfo{title}{Deep learning models for
  forecasting dengue fever based on climate data in {Vietnam}}}.
\newblock {\emph{\JournalTitle{PLOS Neglected Tropical Diseases}}}
  \textbf{\bibinfo{volume}{16}}, \bibinfo{pages}{e0010509},
  \url{10.1371/journal.pntd.0010509} (\bibinfo{year}{2022}).

\bibitem{johansson_evaluating_2016}
\bibinfo{author}{Johansson, M.~A.}, \bibinfo{author}{Reich, N.~G.},
  \bibinfo{author}{Hota, A.}, \bibinfo{author}{Brownstein, J.~S.} \&
  \bibinfo{author}{Santillana, M.}
\newblock \bibinfo{journal}{\bibinfo{title}{Evaluating the performance of
  infectious disease forecasts: {A} comparison of climate-driven and seasonal
  dengue forecasts for {Mexico}}}.
\newblock {\emph{\JournalTitle{Sci Rep}}} \textbf{\bibinfo{volume}{6}},
  \bibinfo{pages}{33707}, \url{10.1038/srep33707} (\bibinfo{year}{2016}).

\bibitem{withanage_forecasting_2018}
\bibinfo{author}{Withanage, G.~P.}, \bibinfo{author}{Viswakula, S.~D.},
  \bibinfo{author}{Nilmini Silva~Gunawardena, Y.~I.} \&
  \bibinfo{author}{Hapugoda, M.~D.}
\newblock \bibinfo{journal}{\bibinfo{title}{A forecasting model for dengue
  incidence in the {District} of {Gampaha}, {Sri} {Lanka}.}}
\newblock {\emph{\JournalTitle{Parasit Vectors}}}
  \textbf{\bibinfo{volume}{11}}, \bibinfo{pages}{262},
  \url{10.1186/s13071-018-2828-2} (\bibinfo{year}{2018}).
\newblock \bibinfo{note}{Place: England}.

\bibitem{polwiang_time_2020}
\bibinfo{author}{Polwiang, S.}
\newblock \bibinfo{journal}{\bibinfo{title}{The time series seasonal patterns
  of dengue fever and associated weather variables in {Bangkok} (2003-2017).}}
\newblock {\emph{\JournalTitle{BMC Infect Dis}}} \textbf{\bibinfo{volume}{20}},
  \bibinfo{pages}{208}, \url{10.1186/s12879-020-4902-6} (\bibinfo{year}{2020}).
\newblock \bibinfo{note}{Place: England}.

\bibitem{sylvestre_data-driven_2022}
\bibinfo{author}{Sylvestre, E.} \emph{et~al.}
\newblock \bibinfo{journal}{\bibinfo{title}{Data-driven methods for dengue
  prediction and surveillance using real-world and {Big} {Data}: {A} systematic
  review.}}
\newblock {\emph{\JournalTitle{PLoS Negl Trop Dis}}}
  \textbf{\bibinfo{volume}{16}}, \bibinfo{pages}{e0010056},
  \url{10.1371/journal.pntd.0010056} (\bibinfo{year}{2022}).
\newblock \bibinfo{note}{Place: United States}.

\bibitem{shashvat_ensemble_2019}
\bibinfo{author}{Shashvat, K.}, \bibinfo{author}{Basu, R.},
  \bibinfo{author}{Bhondekar, P.~A.} \& \bibinfo{author}{Kaur, A.}
\newblock \bibinfo{journal}{\bibinfo{title}{An ensemble model for forecasting
  infectious diseases in {India}.}}
\newblock {\emph{\JournalTitle{Trop Biomed}}} \textbf{\bibinfo{volume}{36}},
  \bibinfo{pages}{822--832} (\bibinfo{year}{2019}).
\newblock \bibinfo{note}{Place: Malaysia}.

\bibitem{reich_challenges_2016}
\bibinfo{author}{Reich, N.~G.} \emph{et~al.}
\newblock \bibinfo{journal}{\bibinfo{title}{Challenges in {Real}-{Time}
  {Prediction} of {Infectious} {Disease}: {A} {Case} {Study} of {Dengue} in
  {Thailand}.}}
\newblock {\emph{\JournalTitle{PLoS Negl Trop Dis}}}
  \textbf{\bibinfo{volume}{10}}, \bibinfo{pages}{e0004761},
  \url{10.1371/journal.pntd.0004761} (\bibinfo{year}{2016}).
\newblock \bibinfo{note}{Place: United States}.

\bibitem{Johansson_Reich_Hota_Brownstein_Santillana_2016}
\bibinfo{author}{Johansson, M.~A.}, \bibinfo{author}{Reich, N.~G.},
  \bibinfo{author}{Hota, A.}, \bibinfo{author}{Brownstein, J.~S.} \&
  \bibinfo{author}{Santillana, M.}
\newblock \bibinfo{journal}{\bibinfo{title}{Evaluating the performance of
  infectious disease forecasts: A comparison of climate-driven and seasonal
  dengue forecasts for mexico}}.
\newblock {\emph{\JournalTitle{Scientific reports}}}
  \textbf{\bibinfo{volume}{6}}, \bibinfo{pages}{33707} (\bibinfo{year}{2016}).

\bibitem{Mosqlimate-project}
\bibinfo{author}{Coelho, F.~C.} \& \bibinfo{author}{Mosqlimate-team}.
\newblock \bibinfo{title}{Mosqlimate project code repository}.
\newblock \bibinfo{howpublished}{\url{https://github.com/Mosqlimate-project}}
  (\bibinfo{year}{2023}).

\bibitem{Vacaro_Coelho_Araujo_Loch_Bot_Pimpale_2024}
\bibinfo{author}{Vacaro, L.~B.}, \bibinfo{author}{Araujo, E.~C.},
  \bibinfo{author}{Loch, S.}, \bibinfo{author}{Pimpale, R.} \&
  \bibinfo{author}{Coelho, F.~C.}
\newblock \bibinfo{title}{Mosqlimate-project/data-platform: 1.1.0},
  \url{10.5281/zenodo.12744275} (\bibinfo{year}{2024}).

\bibitem{Mosqlimate-docs}
\bibinfo{author}{Coelho, F.~C.} \& \bibinfo{author}{Mosqlimate-team}.
\newblock \bibinfo{title}{Data platform documentation}.
\newblock \bibinfo{howpublished}{\url{https://api.mosqlimate.org/docs/}}
  (\bibinfo{year}{2023}).

\bibitem{OpenAPI-Initiative}
\bibinfo{author}{OpenAPI-Initiative}.
\newblock \bibinfo{title}{Openapi specification v3.1.0 | introduction,
  definitions, \& more} (\bibinfo{year}{2021}).

\bibitem{Chaves_Pascual_2007}
\bibinfo{author}{Chaves, L.~F.} \& \bibinfo{author}{Pascual, M.}
\newblock \bibinfo{journal}{\bibinfo{title}{Comparing models for early warning
  systems of neglected tropical diseases}}.
\newblock {\emph{\JournalTitle{PLoS Neglected Tropical Diseases}}}
  \textbf{\bibinfo{volume}{1}}, \bibinfo{pages}{e33} (\bibinfo{year}{2007}).

\bibitem{Fu_Chen_Dong_Luo_Miao_Song_Huang_Sun_2019}
\bibinfo{author}{Fu, T.} \emph{et~al.}
\newblock \bibinfo{journal}{\bibinfo{title}{Development and comparison of
  forecast models of hand-foot-mouth disease with meteorological factors}}.
\newblock {\emph{\JournalTitle{Scientific Reports}}}
  \textbf{\bibinfo{volume}{9}}, \bibinfo{pages}{15691} (\bibinfo{year}{2019}).

\bibitem{Gneiting_Raftery_2007}
\bibinfo{author}{Gneiting, T.} \& \bibinfo{author}{Raftery, A.~E.}
\newblock \bibinfo{journal}{\bibinfo{title}{Strictly proper scoring rules,
  prediction, and estimation}}.
\newblock {\emph{\JournalTitle{Journal of the American Statistical
  Association}}} \textbf{\bibinfo{volume}{102}}, \bibinfo{pages}{359–378},
  \url{10.1198/016214506000001437} (\bibinfo{year}{2007}).

\bibitem{johansson2019open}
\bibinfo{author}{Johansson, M.~A.} \emph{et~al.}
\newblock \bibinfo{journal}{\bibinfo{title}{An open challenge to advance
  probabilistic forecasting for dengue epidemics}}.
\newblock {\emph{\JournalTitle{Proceedings of the National Academy of
  Sciences}}} \textbf{\bibinfo{volume}{116}}, \bibinfo{pages}{24268--24274}
  (\bibinfo{year}{2019}).

\bibitem{bracher2021evaluating}
\bibinfo{author}{Bracher, J.}, \bibinfo{author}{Ray, E.~L.},
  \bibinfo{author}{Gneiting, T.} \& \bibinfo{author}{Reich, N.~G.}
\newblock \bibinfo{journal}{\bibinfo{title}{Evaluating epidemic forecasts in an
  interval format}}.
\newblock {\emph{\JournalTitle{PLoS computational biology}}}
  \textbf{\bibinfo{volume}{17}}, \bibinfo{pages}{e1008618}
  (\bibinfo{year}{2021}).

\bibitem{zanetta_scoringrules_2024}
\bibinfo{author}{Zanetta, F.} \& \bibinfo{author}{Allen, S.}
\newblock \bibinfo{title}{Scoringrules: a python library for probabilistic
  forecast evaluation} (\bibinfo{year}{2024}).

\bibitem{codeco2018infodengue}
\bibinfo{author}{Codeco, C.} \emph{et~al.}
\newblock \bibinfo{journal}{\bibinfo{title}{Infodengue: A nowcasting system for
  the surveillance of arboviruses in brazil}}.
\newblock {\emph{\JournalTitle{Revue d'{\'E}pid{\'e}miologie et de Sant{\'e}
  Publique}}} \textbf{\bibinfo{volume}{66}}, \bibinfo{pages}{S386}
  (\bibinfo{year}{2018}).

\bibitem{araujo2024large}
\bibinfo{author}{Araujo, E.~C.} \emph{et~al.}
\newblock \bibinfo{journal}{\bibinfo{title}{Large-scale epidemiological
  modeling: Scanning for mosquito-borne diseases spatio-temporal patterns in
  brazil}}.
\newblock {\emph{\JournalTitle{arXiv preprint arXiv:2407.21286}}}
  (\bibinfo{year}{2024}).

\bibitem{hersbach2018era5}
\bibinfo{author}{Hersbach, H.} \emph{et~al.}
\newblock \bibinfo{journal}{\bibinfo{title}{Era5 hourly data on single levels
  from 1940 to present}}.
\newblock {\emph{\JournalTitle{Copernicus climate change service (c3s) climate
  data store (cds)}}} \textbf{\bibinfo{volume}{10}} (\bibinfo{year}{2018}).

\end{thebibliography}


\section*{Acknowledgements}

The authors would like to thank the Wellcome Trust for funding Mosqlimate project  through grant number 226088/Z/22/Z


\section*{Author contributions statement}

F.C.C. Conceived and supervised the research.  L.B.V, L.D.A and E.C.A, developed the software, L.B and L.M.C proceeded the analysis analyzed the results I.A A.M.S and F.G validated the epidemiological, climate and other dataset variables and analyzed the results. All authors contributed to writing and reviewing the manuscript.

\section*{Competing interests}

The authors declare no competing interests 

\section*{Figures \& Tables}


\begin{table}
    \centering
    \caption{Metadata fields that are provided on model registration}
    \label{tab:model_metadata}
    \begin{tabular}{l|c|c}
    \hline
        \textbf{Field} & \textbf{Description} & \textbf{Type} \\\hline
        name & Model name &  String \\
        description & Short text describing the model  & Text \\
        repository & URL source-code repository & String \\
        implementation\_language & Language used to generate Predictions & String (selected from fixed list) \\
        disease & Model disease & 'dengue', 'zika' or 'chikungunya' \\
        temporal & Timed-series model & Boolean \\
        spatial & Geo-spatial model & Boolean \\
        categorical & Categorical or Quantitative model & Boolean \\
        ADM\_level & Geographical scope & 0, 1, 2 or 3 \\
        time\_resolution & Periodicity of predictions & Day, week, month, or year \\
        sprint & Model for Sprint 2024/25 & Boolean \\\hline
    \end{tabular}
\end{table}

\begin{table}
    \centering
    \caption{Prediction metadata.}
    \label{tab:pred_metadata}
    \begin{tabular}{l|c|c}
    \hline
        \textbf{Field} & \textbf{Description} & \textbf{Type}\\\hline
        model & ID of the Prediction's Model & Integer \\
        description & Prediction's description & Text \\
        commit & Git commit hash (model version) & String \\
        predict\_date & Date when Prediction was generated & Date (YYYY-mm-dd) \\
        prediction & Model output & JSON object\\
        \hline
    \end{tabular}
\end{table}

\begin{table}
    \centering
    \caption{Prediction columns required in the JSON object.}
    \label{tab:pred_table_metadata}
    \begin{tabular}{l|c|c}
    \hline
        \textbf{Column name} & \textbf{Description} & \textbf{Type}\\\hline
        date& Date of the predicted value & Date (YYYY-mm-dd) \\
        pred& Median predicted value& float \\
        lower & Lower predicted value of the confidence interval& float\\
        upper & Upper predicted value of the confidence interval& float \\
        adm\_0 & Geocode associated with the ADM 0 region (national) & Integer/String\\
        adm\_1 & Geocode or UF associated with the ADM 1 region (state) & Interger/String\\
        adm\_2 & Geocode associated with the ADM 2 region (municipality) & Integer\\
        adm\_3 & Geocode associated with the ADM 3 region (sub-municipality) & Integer\\
        \hline
    \end{tabular}
\end{table}

\begin{table}[h]
    \centering
    \caption{Main variables of the Infodengue dataset and their descriptions.}
    \begin{tabular}{l|p{9cm}}
        \textbf{Variable name}&\textbf{Description} \\\hline
        data\_iniSE & Start date of epidemiological week   \\
        SE 	& 	Epidemiological week\\
        casos 	& 	Number of notified cases per week\\
        casos\_est 	& 	Estimated number of cases per week by the nowcasting model\\
        casos\_prov 	& 	Probable number of cases per week (notified cases - discarded cases)\\
        municipio\_geocodigo 	& 	IBGE's municipality code\\
        p-rt1 	& 	Probability (Rt > 1)\\
        p\_inc100k 	& 	Estimated incidence rate (cases per pop x 100K)\\
        nivel 	& 	Alert level (1 = green, 2 = yellow, 3 = orange, 4 = red)\\
        versao\_modelo 	& 	Alert Model version\\
        Rt 	& 	Point estimate of the effective reproduction number of cases\\
        municipio-nome 	& 	Municipality's name\\
        pop 	& 	Population (IBGE)\\
        receptivo 	& 	Indicates climate receptivity, i.e., conditions for high vectorial capacity. 0 = unfavorable, 1 = favorable, 2 = favorable this week and last week, 3 = favorable for at least three weeks\\
        transmissao 	& 	Evidence of sustained transmission: 0 = no evidence, 1 = possible, 2 = likely, 3 = highly likely\\
        nivel\_inc 	& 	Estimated incidence below pre-epidemic threshold, 1 = above pre-epidemic threshold but below epidemic threshold, 2 = above epidemic threshold\\
        \hline
    \end{tabular}
    \label{tab:infodengue}
\end{table}

\begin{table}[h]
    \centering
    \caption{Data dictionary for the Episcanner dataset.}
    \begin{tabular}{c|c|p{9cm}}
         \textbf{Parameter name} 	&\textbf{Type}& 	\textbf{Description}\\\hline
        disease 	&str 	&dengue, zika or chik\\
        CID10 	&str 	&Disease's ICD code\\
        year 	&int 	&Year analyzed\\
        geocode 	&int 	&City's geocode\\
        muni-name 	&str 	&City's name\\
        peak-week 	&float 	&Estimated epidemiological week of the peak of the epidemic\\
        beta 	&float &	Transmissibility rate\\
        gamma 	&float 	&Recovery rate\\
        R0 	&float& 	Basic reproduction number\\
        total-cases 	&int 	&Annual total number of cases\\
        alpha 	&float 	&Parameter of the Richard's model\\
        sum-res 	&float 	&Sum of residuals\\
        ep-ini 	&str 	&Estimated starting week of the epidemic. Format: YYYYWW\\
        ep-end 	&str &	Estimated Ending week of the epidemic. Format: YYYYWW\\
        ep-dur 	&int 	&Estimated duration of the epidemic in weeks\\
        \hline
    \end{tabular}
    
    \label{tab:episcanner-dict}
\end{table}

\begin{table}
    \centering
    \caption{Available output items in the climate dataset.}
    \label{tab:clim_vars}
    \begin{tabular}{l|l|c}
    \hline
       \textbf{Variable}  & \textbf{Description} & \textbf{Unit}\\\hline
        date & Date in the format YYYY-mm-dd, representing the day of the year. & Day of the year \\
        geocodigo & IBGE's municipality code. & int \\
        temp\_min & Minimum daily temperature. & $^oC$ \\
        temp\_med & Average daily temperature. & $^oC$ \\
        temp\_max & Maximum daily temperature. & $^oC$ \\
        precip\_min & Minimum daily precipitation. & mm \\
        precip\_med & Average daily precipitation. & mm \\
        precip\_max & Maximum daily precipitation. & mm \\
        precip\_tot & Total daily precipitation. & mm \\
        pressao\_min & Minimum daily sea level pressure. & Atm \\
        pressao\_med & Average daily sea level pressure. & Atm \\
        pressao\_max & Maximum daily sea level pressure. & Atm \\
        umid\_min & Minimum daily relative humidity. & \% \\
        umid\_med & Average daily relative humidity. & \% \\
        umid\_max & Maximum daily relative humidity. & \% \\
        \hline
    \end{tabular}
\end{table}


\end{document}